\documentstyle[11pt,epsfig]{article}
\newcommand{\be}{\begin{equation}}
\newcommand{\ee}{\end{equation}}
\newcommand{\beeq}{\begin{eqnarray}}
\newcommand{\eeeq}{\end{eqnarray}}

\def\kbo{{\bf k}}

\def\lbo{{\bf l}}

\def\rbo{{\bf r}}

\def\Qdash{\overline{Q}}

\def\gev{\mbox{\rm GeV}}
\newlength{\dinwidth}
\newlength{\dinmargin}
\begin{document}
\titlepage
\begin{flushright}
DESY 02--037\\
March 2002
\end{flushright}

\vspace*{1in}
\begin{center}
{\Large \bf A Modification of the Saturation Model: DGLAP Evolution}\\
\vspace*{0.5in}
J. \ Bartels$^{(a)}$, 
K. \ Golec-Biernat$^{(a,b)}$
and H.\ Kowalski$^{(c)}$ \\
\vspace*{0.5cm}
{\it $^{(a)}$II.\ Institut f\"ur Theoretische Physik, 
    Universit\"at Hamburg,\\ Luruper Chaussee 149, 
    D-22761 Hamburg, Germany\\
\vskip 2mm
$^{(b)}$H. Niewodnicza\'nski Institute of Nuclear Physics,
   Krak\'ow, Poland} \\
\vskip 2mm
$^{(c)}$ {\it Deutsches Electronen Synchrotron DESY, Hamburg, Germany} \\
\end{center}
\vspace*{2cm}

\vskip1cm 
\begin{abstract}
\noindent
We propose to modify the saturation model  of
Golec-Biernat and W\"usthoff by including DGLAP evolution.
We find considerable improvement for the total deep inelastic cross section, 
in particular in the large $Q^2$ region. The successful description of 
DIS diffraction is preserved.  
\end{abstract}
\newpage

The saturation model ~\cite{GBW1,GBW2,GBW3} has provided a successful
description of HERA deep inelastic scattering data, in particular for the 
transition from the perturbative region to the nonperturbative
photoproduction region. This includes both the total $\gamma^* p$ cross
section and the DIS diffractive cross section. Whereas the formulae
are particularly appealing through their simplicity, they also have
an attractive theoretical background, namely the idea of saturation.
Despite its success, the model suffers from shortcomings which should
be cured. In particular, the model does not include logarithmic scaling 
violations,
i.e. at larger values of $Q^2$ it does not exactly match with QCD evolution 
(DGLAP). This becomes clearly visible in the energy dependence of 
$\sigma_{tot}^{\gamma^*p}$ in the region $Q^2>20$ GeV$^2$ where the
model predictions are below the data. One expects that QCD evolution
should enhance the cross section in this region.

It is the purpose of this letter to
propose a modification of the saturation model. We attempt to preserve
the success of the model in the low-$Q^2$ and in the transition region,
while incorporating DGLAP evolution in the large-$Q^2$ domain.
Since the energy dependence in the large-$Q^2$
region is mainly due to the behaviour of the dipole cross section at small 
dipole sizes $r$, our changes will affect mostly the small-$r$ region.
At the same time, particular attention will be given to DIS diffraction 
for which the saturation model correctly describes the energy dependence.
Since the inclusive diffractive cross section mostly depends upon the
large-$r$ behaviour of the dipole cross section, we attempt to leave  
the dipole cross section unchanged in this region. A recent attempt
\cite{LL} along the same lines indicates that, in fact, diffraction
provides a highly nontrivial restriction on possible modifications of the
saturation model.

\section{The Model}
\label{sec:1}
Before we describe the modifications of the saturation model,
we briefly review the main features of its original formulation.
Within the dipole formulation of the $\gamma^*p$ scattering,
\be
\label{eq:totalcross}
\sigma_{T,L}^{\gamma^*p}(x,Q^2)\,=\,
\int d^2r\int dz\; \psi^*_{T,L}(Q,r,z)\; \hat\sigma (x,r)\;
\psi_{T,L}(Q,r,z), 
\ee
where $T,L$ denotes the virtual photon polarisation,
the dipole cross section was proposed to have the form
\be
\label{eq:sighat}
\hat\sigma (x,r)\,=\,\sigma_0\,\left\{
1\,-\,\exp\left(-\frac{r^2}{4 R_0^2(x)}\right)
\right\}\,,
\ee
where $R_0(x)$ is the saturation scale which decreases when $x\rightarrow 0$,
\be
\label{eq:satscale}
R_0^2(x)\,=\,\frac{1}{\mbox{\rm GeV}^2}\,
\left(\frac{x}{x_0}\right)^\lambda\,.
\ee
In order to be able to study the formal photoproduction limit,
the Bjorken variable $x=x_{B}$ was modified to be
\be
\label{eq:bjv}
x\,=\,x_{B}\left(1+\frac{4m_q^2}{Q^2}\right)\,=\,\frac{Q^2+4m_q^2}{W^2}\,,
\ee
where $m_q$ is an effective quark mass, and $W^2$ denotes the $\gamma^* p$
center-of-mass energy squared.
The parameters  of the model,
$\sigma_0=23~{\rm mb}$, $\lambda=0.29$ and
$x_0=3\cdot 10^{-4}$ (for the assumed quark mass
$m_q=140~\mbox{\rm{MeV}}$) were found from a fit to small-$x$ data \cite{GBW1}.
For alternative forms of  the dipole cross section
parameterisation see \cite{OTHERS}.

As it is well known \cite{DIPGLUE}, in the small-$r$ region the dipole cross section
is related to the gluon density
\be
\label{eq:smallr}
\hat\sigma(x,r) \,\simeq\,
\frac{\pi^2}{3}\,r^2\,\alpha_s\,xg(x,\mu^2)\,.
\ee
where the scale $\mu^2$ for small $r$ behaves as $C/r^2$. Equation (5)
is valid in the double logarithmic approximation in which the constant $C$
is not determined. In the saturation model, eq.~(\ref{eq:sighat}),
we find for small $r\ll 2 R_0(x)$ 
\be
\hat\sigma (x,r)\,\simeq  \,\frac{\sigma_0\,r^2}{4\, R_0^2(x)}\,,
\ee
i.e., the gluon density is modelled as:
\be
xg(x,\mu^2)\,=\,\frac{3}{4\pi^2\alpha_s}\,\frac{\sigma_0}{ R_0^2(x)}\,.
\ee
For fixed $\alpha_s$ this gluon density is clearly scale independent,
which contradicts the QCD DGLAP evolution.
Thus, in order to correctly take into account the  scale dependence
as given by the  DGLAP
evolution equations we have to modify the small-$r$ behaviour of the dipole 
cross section by incorporating the properly evolved gluon density.
At the same time, we wish to preserve the idea of saturation,
which reflects unitarity, and to keep unaltered the large-$r$ behaviour of the dipole
cross section which determines the diffractive cross section.

Therefore, we propose the following modification of the model (\ref{eq:sighat})
\be
\label{eq:sighatnew}
\hat\sigma (x,r)\,=\,\sigma_0\,\left\{
1\,-\,\exp\left(-\frac{\pi^2\,r^2\,\alpha_s(\mu^2)\,xg(x,\mu^2)}
{3\,\sigma_0}\right) \right\}\,,
\ee
where the scale $\mu^2$ is assumed to have the form
\be
\label{eq:scale}
\mu^{2}\,=\,\frac{C}{r^2}\,+\,\mu_0^2\,.
\ee
The parameters $C$ and $\mu_0^2$ will be determined from a fit
to DIS data. In a first approximation,
$g(x,\mu^2)$ is evolved with the leading order DGLAP
evolution equation for the gluon density. In the spirit of the
small-$x$ limit, we neglect  quarks in the evolution equations.
We assume the following gluon
density at the initial scale $Q_0^2=1~\mbox{\rm GeV}^2$:
\be
\label{eq:gluon}
xg(x,Q_0^2)\,=\,A_g\,x^{-\lambda_g}\,(1-x)^{5.6}\,,
\ee
where $A_g$ and $\lambda_g$ are parameters to be determined from a fit to
data. The exponent $5.6$, determining the
large $x$ behaviour, is motivated by one of the versions of
the MRST parameterisation \cite{MRST01} of the gluon density.

For small $r$, the exponential in  (\ref{eq:sighatnew})
can be expanded in powers  of its argument, and  the relation 
(\ref{eq:smallr}), with the running $\alpha_s=\alpha_s(\mu^2)$, is found.
In contrast to the original dipole cross section,
the rise in $1/x$ now has become $r$-dependent. When inserting $\hat{\sigma}$
into (\ref{eq:totalcross}) and convoluting with the 
photon wave function, the integrand peaks near $r \sim 2/Q$ for large
$Q^2$, and the argument
of the gluon density turns into $\mu^2 \approx Q^2$. Consequently,
with increasing $Q^2$, DGLAP evolution will strengthen the rise in $1/x$,
whereas in the original saturation model the power of $1/x$ had been constant.
For sufficiently large $r$, the scale $\mu^2$ is frozen
at the value $\mu_0^2$. This prevents the effective scale of the gluon density
from becoming unreasonably small. The saturation value of the dipole cross 
section is $\hat\sigma(x,r)\,\approx\,\sigma_0$,
as in the original model (\ref{eq:sighat}). The transition 
from the small to the large $r$ region depends on $x$, but 
in detail it will be different from the original model.  This will be discussed
in detail in the section presenting numerical results.

\section{Momentum space formulation}

Although in this letter we will restrict ourselves to the total (and later
on to the diffractive) cross section, it is instructive to rephrase these
features in momentum space. In a future step, we intend to study the
effects of saturation in more exclusive final states, and the translation of
our dipole cross section into momentum space may serve as a first step
into this direction. For this purpose let us start with the
$k_T$-fatorization formula \cite{KTFAC} for the $\gamma^*p$ cross section,
e.g for the transversely polarised photon \cite{NIKO},
\be
\label{eq:cst}
\sigma_T^{\gamma*p}
\,=\,
\frac{\alpha_{em}}{\pi} \sum_{f} e_f^2\,
\int_0^1 dz\,
[z^2+(1-z)^2]\,
\int d^2\kbo\,
\int \frac{d^2\lbo}{l^4}\ \alpha_s f(x,l^2)\,
\left\{\frac{\kbo}{\kbo^2+\Qdash^2}
-\frac{\kbo+\lbo}{(\kbo+\lbo)^2+\Qdash^2} \right\}^2\,,
\ee
where $f(x,l^2)$ is the gluon amplitude describing an interaction of the
$q\bar{q}$ pair with the proton, $\lbo$ is the transverse momentum
of the gluon coupled to the quark pair  and $\overline{Q}^2=z(1-z)Q^2$. 
Using the relation
\be
\frac{\kbo}{{\bf{k}}^2+\overline{Q}^2}\,=\,
i\,\overline{Q}\,
\int\frac{d^2\rbo}{2\pi}\,
e^{i \kbo\cdot \rbo}\,\frac{\rbo}{r}\,K_1(\overline{Q} r)\,,
\ee
the following formula is found
\beeq
\label{eq:pesh}
\sigma_T^{\gamma*p}
\nonumber
&=&
\frac{3\alpha_{em}}{2\pi^2} \, \sum_{f} e_f^2  \int_0^1 dz\ 
[z^2+(1-z)^2]\;\overline{Q}^2
\int {d^2\rbo} \int {d^2\rbo^\prime}\,
 \frac{\rbo\cdot\rbo^\prime}{r\, r^\prime}\,K_1(\overline{Q} r)\,
K_1(\overline{Q} r^\prime)\,
\\  \nonumber
\\
&\times&
\int \frac{d^2\kbo}{(2\pi)^2}\, e^{i \kbo \cdot (\rbo-\rbo^\prime)}\,D(\rbo,\rbo^\prime)
\eeeq
where
\be
\label{eq:d}
D(\rbo,\rbo^\prime)\,=\,\frac{2\pi}{3}\
\int \frac{d^2\lbo}{l^4}\ \alpha_s f(x,l^2)\ (1-e^{i\lbo\cdot \rbo})\
(1-e^{-i\lbo\cdot \rbo^\prime})\,.
\ee
If  the argument of the strong coupling  $\alpha_s$ and the variable $x$
in  the gluon amplitude in $D(\rbo,\rbo^\prime)$
do not depend on the quark transverse momenta $\kbo$, 
the integration over $\kbo$ in eq.~(\ref{eq:pesh})  
gives the delta function  $\delta^2(\rbo-\rbo^\prime)$
which reflects the conservation of a dipole transverse size vector $\rbo$
during the collision. In this case the dipole formula (\ref{eq:totalcross}) is obtained
with the following identification of the dipole cross section
\be
\label{eq:1}
\hat\sigma(x,\rbo)
\,=\,
\frac{2\pi}{3}\
\int \frac{d^2\lbo}{l^4}\ \alpha_s f(x,l^2)\ (1-e^{i\lbo\cdot \rbo})\
(1-e^{-i\lbo\cdot \rbo})\,.
\ee
Going beyond the  leading $\log(1/x)$ approximation  
in which eq.~(\ref{eq:cst}) was derived,
e.g. by taking into account the exact gluon kinematics
\cite{PESH} or considering a quark virtuality $\kbo^2+\overline{Q}^2$ as an argument
of the running coupling $\alpha_s$, we find that
$\rbo$ is no longer conserved during  the scattering  process, and
the simple relation (\ref{eq:1})  ceases to exist. As a result, the
$k_T$-factorization formula (\ref{eq:cst}) can no longer be written in
the form  (\ref{eq:totalcross}), and the simple dipole picture fails.
We want to avoid this situation, thus   we assume that the argument of $\alpha_s$ is given
by the gluon momentum $\lbo^2$ and  $x=x_{Bj}$.  Since the integration in (\ref{eq:1})
includes also  small momenta, the modelling of the infrared behaviour of $\alpha_s$ cannot
be avoided. However, we will hide this fact by analysing the combined
quantity $\alpha_s f(x,l^2)$.

From the requirement that in the double logarithmic limit (DLL) formula
(\ref{eq:cst}) should be consistent with the DLL of the DGLAP formalism
one can derive a relation between
the gluon amplitude $f(x,l^2)$ and the conventional gluon distribution
$xg(x,Q^2)$. Starting from eq.~(\ref{eq:cst}), using the relation
$F_T=Q^2/(4\pi^2\alpha_{em})\,\sigma^{\gamma^*p}_T$  and
imposing the strong ordering condition: $\lbo^2\ll \kbo^2\ll Q^2$,
one arrives at
\be
\frac{\partial F_T(x,Q^2)}{\partial \log Q^2}\,=\,
\frac{1}{3\pi}\,\sum_f e^2_f\,\int^{Q^2} \frac{d^2\lbo}{\pi\, l^2}\,
\alpha_s f(x,l^2)\,.
\ee
By comparison with an analogous formula obtained in the DLL
of the DGLAP evolution equations, one finds the following relation at large $Q^2$
\be
\label{eq:2}
\alpha_s(Q^2)\,xg(x,Q^2)\,=\,\int^{Q^2} \frac{d^2\lbo}{\pi\, l^2}\,
\alpha_s f(x,l^2)\,.
\ee

In the model (\ref{eq:sighatnew}) we  go beyond the $k_T$-factorization formula
(\ref{eq:cst}) where $f(x,l^2)$ represent a two gluon amplitude.
In the region of small $l^2$, the relation (\ref{eq:2}) between
$f(x,l^2)$ and the gluon density no longer holds, and
$f(x,l^2)$ is defined through relation (\ref{eq:1}) where for the
l.h.s we use our model (\ref{eq:sighatnew}). 
In general, provided the dipole cross section has a finite limit:
$\lim_{r\rightarrow \infty}\,\hat\sigma(x,r)\,=\,\hat\sigma_{\infty}(x)$,
the equation (\ref{eq:1}) can be inverted with the help of the following relation:
\beeq
\nonumber
\frac{\alpha_s\,f(x,l^2)}{l^4}
&=&
\frac{3}{4\pi}\,
\int \frac{d^2\rbo}{(2\pi)^2}\,\exp\{i\lbo\cdot \rbo\}\,
\left\{\hat\sigma_{\infty}(x)\,-\,\hat\sigma (x,r)
\right\}=
\\ \nonumber
\\
\label{eq:4}
&=&
\frac{3}{8\pi^2}\,
\int_0^\infty dr\,r\,J_0(lr)\,\left\{\hat\sigma_{\infty}(x)\,-\,\hat\sigma (x,r)
\right\}\,.
\eeeq
In the original dipole model we find \cite{GBW2}:
\be
\label{eq:gludip}
\alpha_s\,f(x,l^2)\,=\,\frac{3\sigma_0}{4\pi^2}\,\,R_0^2(x)\,\,l^4\,
\exp\{-R_0^2(x)\,l^2\}\,.
\ee
For the modified dipole cross section, this inversion has to be done
numerically.  DGLAP evolution will affect mainly the large-$l$ behaviour while
at small $l$ our modification should be less severe.
The most interesting question to be addressed below concerns the transition
region: to what extent does our modification affect the region of moderate momenta,
i.e. could one `see' saturation in diffractive final states ?

\section{Numerical results}

Let us now turn to numerical results.
We performed global fits to the DIS data with $x<0.01$ in the range of
$Q^2$ between $0.1$ and $500~\mbox{\rm GeV}^2$. For H1 and ZEUS HERA 
experiments the new 1996-1997 data sets were used~\cite{ZEUSBPT,ZEUSHIGH,H1}.
In addition to the HERA data also the data of the E665 experiment~\cite{E665}
were used. The statistical and systematic errors were added in quadrature.
The number of degrees of freedom, $N_{df}$, was around 330.

The new data sets are considerably more precise
(with much smaller statistical and systematic errors)
then the  ones used in the
original analysis~\cite{GBW1}. As a preparatory step, we applied the
original model (\ref{eq:sighat}), using the
parameter values of the original fit, to the new data and obtained rather
high value of  $\chi^2/N_{df} \sim 3$ (for the old data, the corresponding
value was $\chi^2/N_{df}=1.18$). Next, we let the new data
to determine their own values of the parameters of
the original model, $\sigma_0$, $\lambda$ and $x_0$ in eq.~(\ref{eq:sighat}).
This led to an
improvement of the fit, $\chi^2/N_{df} \sim 2.2$. Nevertheless, this
relatively poor agreement indicates that the original model is doing not so 
well with the new data, especially for large values of $Q^2$.
As a first step for the
improvement, we modify the dipole cross section at small values of $r$ by including QCD
DGLAP evolution, as given in eq.~(\ref{eq:sighatnew}).

In the modified saturation model,
there are  five parameters to be determined:
$\sigma_0$, $C$, $\mu_0^2$, $A_g$ and $\lambda_g$ from
eqs.~(\ref{eq:sighatnew}, \ref{eq:scale}, \ref{eq:gluon}).
We use  the leading order DGLAP evolution equation for the gluon density,
and we put $\Lambda_{QCD}=200~\mbox{\rm MeV}$  in $\alpha_s$ and set the number of
active flavours $N_f=3$. Thus, although the evolution 
equation for the gluon is decoupled from the quarks, their presence is encoded
in the assumed value of $N_f$.

\begin{table}[t]
\begin{center}
\begin{tabular}{|r|r||r|r|r|r|r||r|}
\hline
&$m_q~(\mbox{\rm MeV})$&~$\sigma_0~(\mbox{\rm mb})$~~&~~$A_g$~~~~&$\lambda_g$~~~~~~&
~$C$~~~~&~~$\mu_0^2$~~~~~&~~~$\chi^2/N_{df}$\\
\hline\hline
{\bf Fit~1}~~&{\bf 140}~~~~~&{\bf 23.0}~~~&1.20~~~~&~~~0.28~~~~&~~~0.26~~~~&
~~~0.52~~~~&1.17~~\\
\hline
{\bf Fit~2~~}&{\bf 0}~~~~~~&{23.8}~~~&~~13.71~~~~&~~~-~0.41~~~~&~~~11.10~~~~&
~~~{\bf 1.00}~~~~&0.97~~\\
\hline
\end{tabular}                    ~~~
\end{center}
\caption{The parameters of the fits to the ZEUS, H1 and E665 data with
$x<0.01$ (333 points). The H1 data was rescaled by a factor of $1.05$. The numbers in
bold are fixed during the fits.}
\label{Table:1}
\end{table}

We performed first the fit leaving all five parameters free and assumed
the value of the light quark mass $m_q=140~\mbox{\rm MeV}$, as in the original
formulation \cite{GBW1}. A good quality fit was
obtained with $\chi^2/N_{df}=1.05$. The found value of the dipole cross section
$\sigma_0=27.4~\mbox{\rm mb}$, however,  was higher  than the saturation
model value, $\sigma_0=23~\mbox{\rm mb}$. Also, the corresponding value of
the photoproduction cross section
($\sigma^{\gamma p}=204~\mbox{\rm mb}$) was significantly higher
than the measured value
$174\pm 1(st) \pm 13(sys)~\mbox{\rm mb}$ at $W=209~\mbox{\rm GeV}$ \cite{ZEUSFOT}.
Thus we decided to decrease $\sigma_0$ and fixed it to the saturation model value
$23~\mbox{\rm mb}$. This value is also advantageous for
the description of the inclusive DIS diffractive cross section which
is more sensitive to large dipole sizes,
i.e., to the saturation region, than the total $\gamma^* p$ cross
section~$\cite{GBW2}$.
The resulting parameters in such  fit are presented in Table~1 ({\bf Fit~1}).
The description is slightly worse than that described above, but both the photoproduction
cross section  ($\sigma^{\gamma p}=189~\mbox{\rm mb}$) and the diffractive cross section
are properly described. This is because we modified only the small dipole size part of
the dipole cross section (\ref{eq:sighat}), without affecting the saturation part.
The found gluon density gives $39 \%$ of the total proton momentum
carried by gluons resolved at the initial scale $Q_0^2=1~\mbox{\rm GeV}^2$.

\begin{figure}[t]
  \vspace*{0.0cm}
     \centerline{
         \epsfig{figure=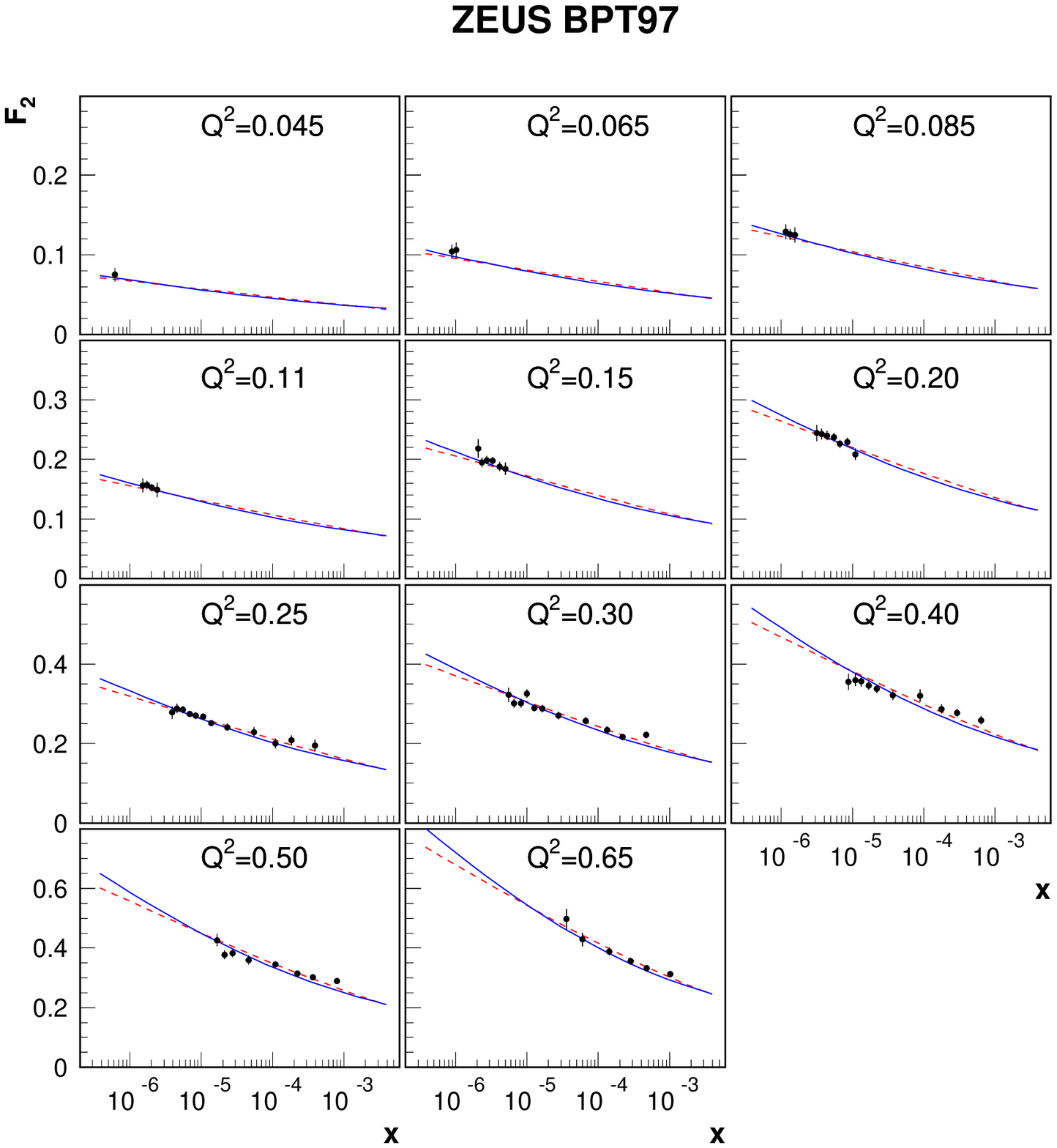,width=17cm}
           }
\vspace*{0.5cm}
\caption{\it $F_2$ as a function of $x$ for fixed low  $Q^2$
values. The comparison with the low $Q^2$ data from ZEUS.
The solid lines: the model with the DGLAP evolution (\ref{eq:sighatnew}) (FIT 1)
and the dotted lines: the saturation model (\ref{eq:sighat}).
\label{fig:3}}
\end{figure}

\begin{figure}[t]
  \vspace*{0.0cm}
     \centerline{
         \epsfig{figure=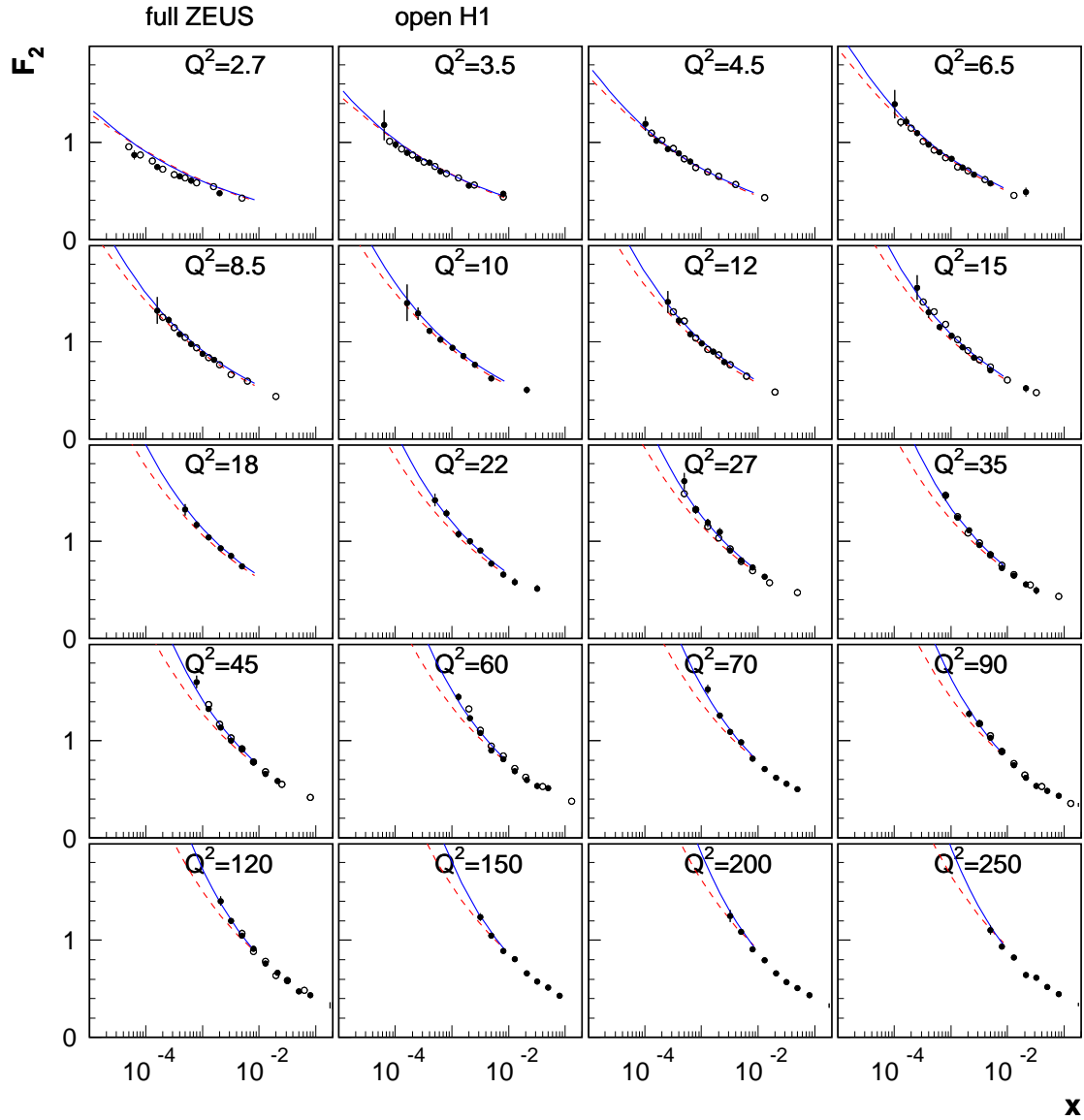,width=17cm}
           }
\vspace*{0.5cm}
\caption{\it H1 and ZEUS data on $F_2$ as a function of $x$
for fixed values of $Q^2>1~\mbox{\rm GeV}^2$ and the saturatiom model curves.
The solid lines: the model with the DGLAP evolution  (\ref{eq:sighatnew}) (FIT 1)
and  the dotted lines: the saturation model (\ref{eq:sighat}).
\label{fig:4}}
\end{figure}

\begin{figure}[t]
  \vspace*{0.0cm}
     \centerline{
         \epsfig{figure=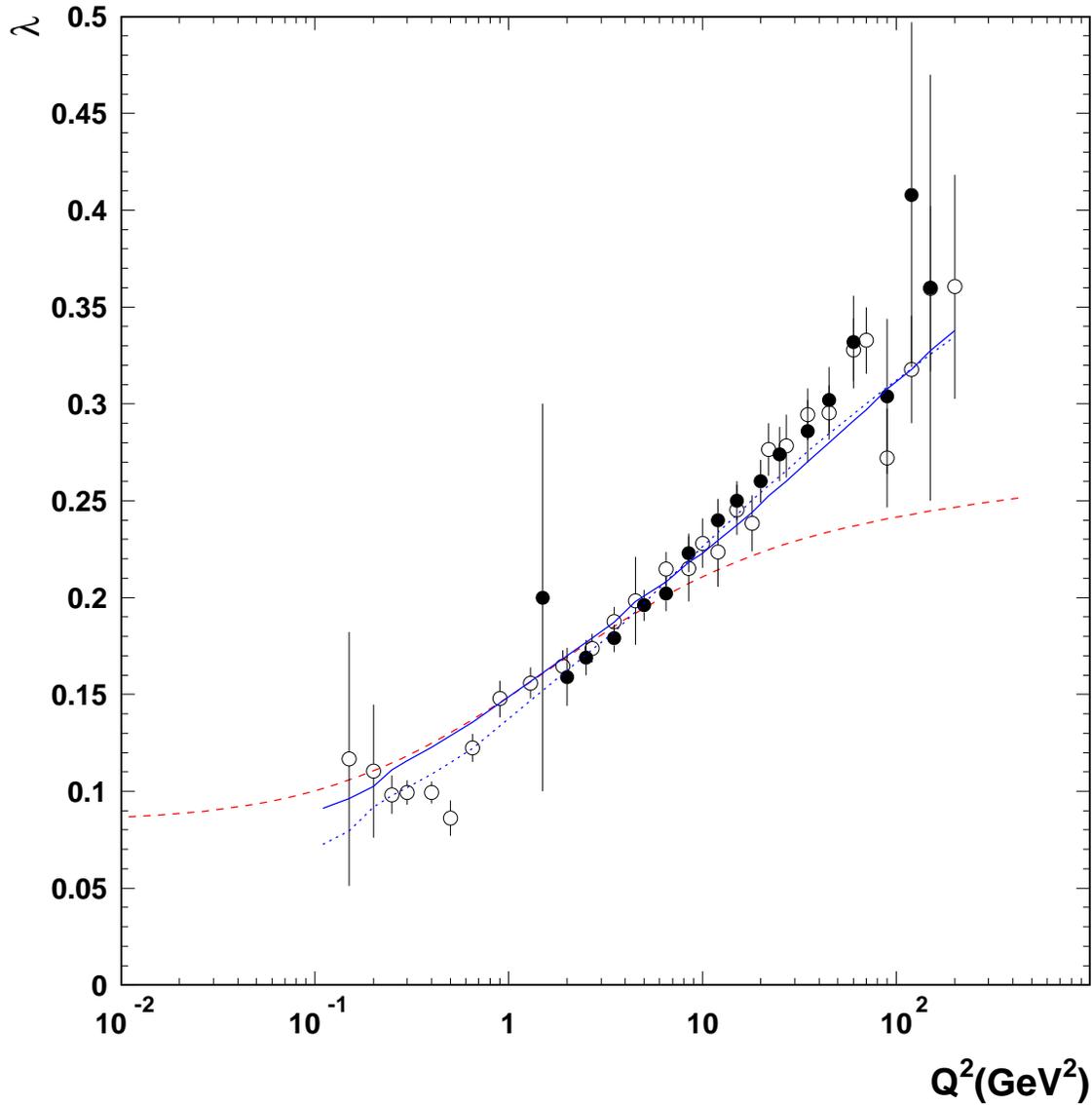,width=17cm}
           }
\vspace*{0.5cm}
\caption{\it The effective slope $\lambda(Q^2)$
from the parameterization
$F_2\sim x^{-\lambda(Q^2)}$ as a function of $Q^2$.
The model with the DGLAP evolution (\ref{eq:sighatnew}): the solid line (FIT 1)
and the dotted line  (FIT 2). The  saturation model (\ref{eq:sighat}): the dashed line.
The open circles: ZEUS analysis and
the full circles: H1 data \cite{H1SLOPES}.
\label{fig:6}}
\end{figure}

\begin{figure}[p]
\hspace{0.cm}
\centerline{
\epsfig{file=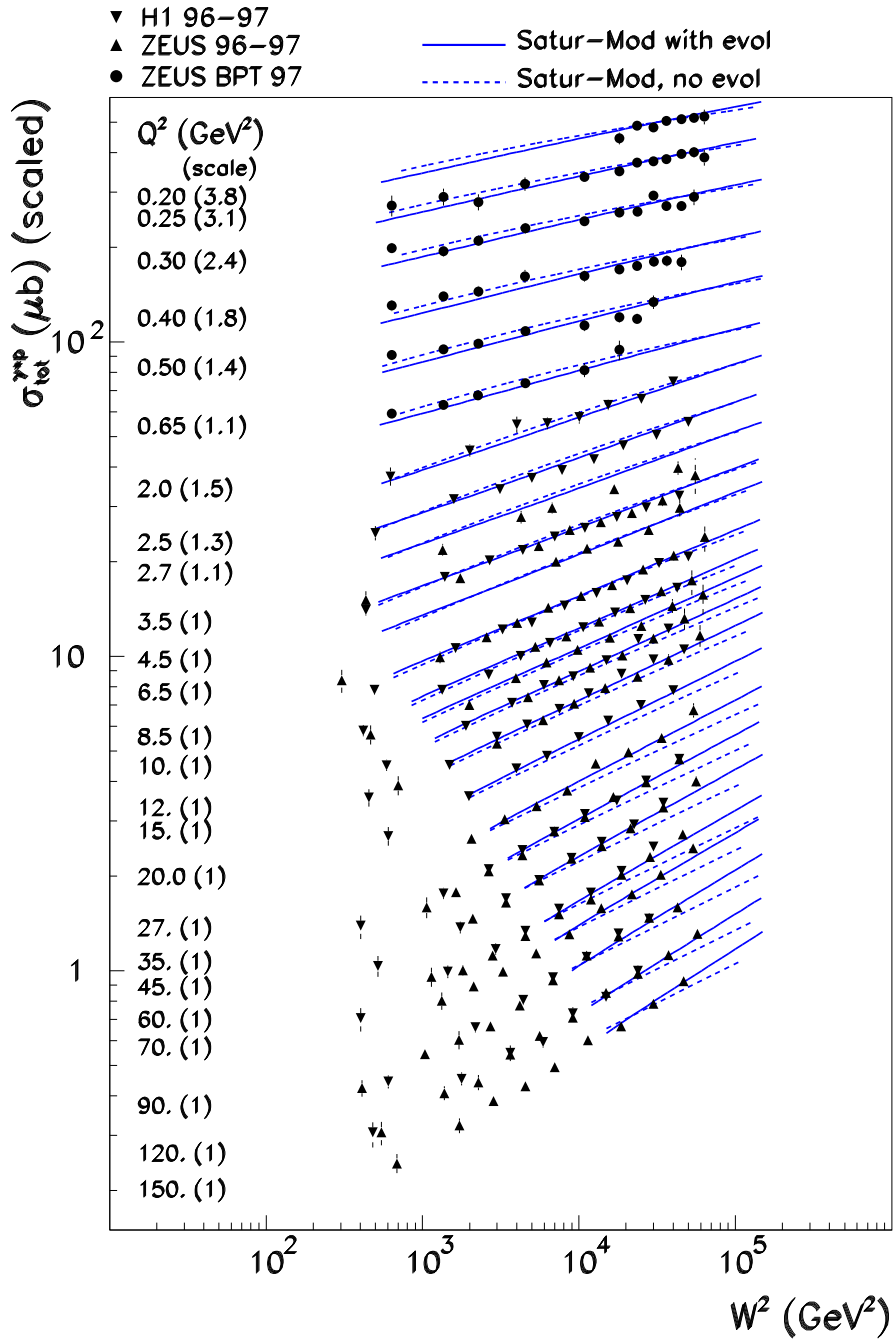,height=19.0cm,width=13.0cm}
}
\caption{\sl  The $\gamma^* p$ cross section as a function of energy $W^2$ at various
$Q^2$. The solid lines: the model with the DGLAP evolution (\ref{eq:sighatnew}) (FIT 1)
and the dotted line: the saturation model (\ref{eq:sighat}), shown for $x<0.01$.
}
\label{fig:sigtotw2}
\end{figure}

\begin{figure}[t]
  \vspace*{-0.5cm}
     \centerline{
         \epsfig{figure=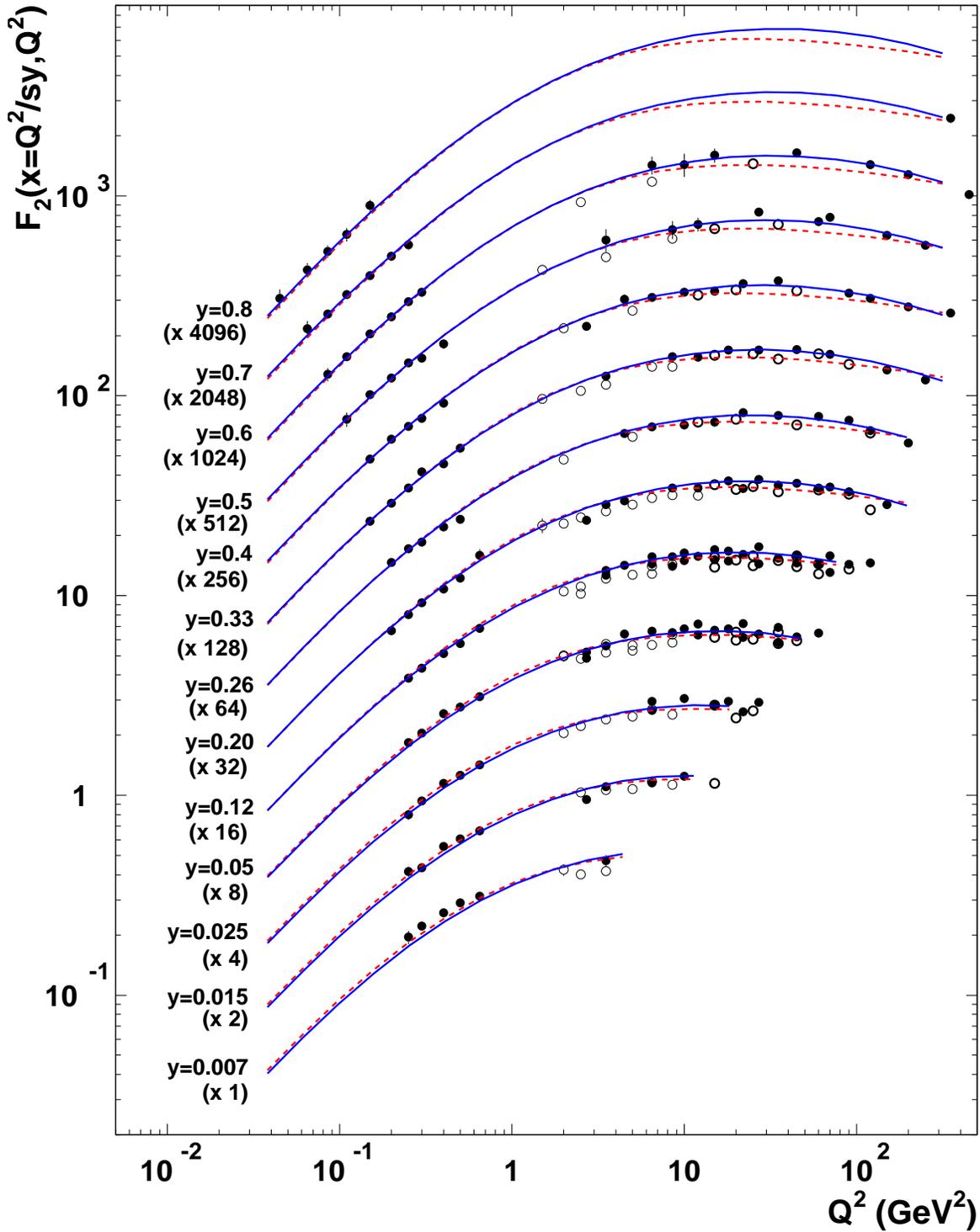,width=17cm}
           }
\vspace*{0.0cm}
\caption{\it $F_2(x,Q^2)$ as a function of $Q^2$ for fixed $y=Q^2/(s x)$.
The solid lines: the model with DGLAP evolution
(\ref{eq:sighatnew}) (FIT 1) and  the dashed lines: the saturation model
(\ref{eq:sighat}).
The curves are plotted for $x<0.01$. Full circles: ZEUS data and open circles: H1 data.
\label{fig:2}}
\end{figure}

The results of { Fit 1} are compared to the data on $F_2$ in Fig.~\ref{fig:3}
for $Q^2<1~\mbox{\rm GeV}^2$ and in  Fig.~\ref{fig:4} for large $Q^2$ points.
In all presented plots, the solid lines  refer to the results obtained with the
DGLAP improved model (\ref{eq:sighatnew}) and  the dashed lines correspond to
the saturation model (\ref{eq:sighat}) with the original parameters from
\cite{GBW1}. We see that the  DGLAP evolution
significantly improves agreement with the data at large $Q^2$ while
at small $Q^2$ the results are practically the same.
This effect is summarised in Fig.~\ref{fig:6} where the effective slopes $\lambda(Q^2)$,
obtained from the  parameterisation of $F_2$ at small $x$: $F_2\sim x^{-\lambda(Q^2)}$,
are plotted.
Thus,  the DGLAP modification of the dipole cross section for small $r$
is crucial for much better agreement with the data.
The same effective slopes characterise the energy dependence of the
$\gamma^* p$ cross section: $\sigma^{\gamma^* p}\sim (W^2)^{\lambda(Q^2)}$.
The change from a soft dependence at small $Q^2$ to  a hard one for
large $Q^2$ is shown in Fig.~\ref{fig:sigtotw2}.   In Fig.~\ref{fig:2} we show
another aspect of the transition of $F_2$
to low $Q^2$ values, namely the emergence of the behaviour: $F_2\sim Q^2$
approached in the limit $Q^2\rightarrow 0$ and $y=W^2/s$ fixed.   The class of
the saturation models
described here nicely reproduce this behaviour, see recent Ref.~\cite{GB1} for more details
on this transition.

\begin{figure}[t]
  \vspace*{0.0cm}
     \centerline{
         \epsfig{figure=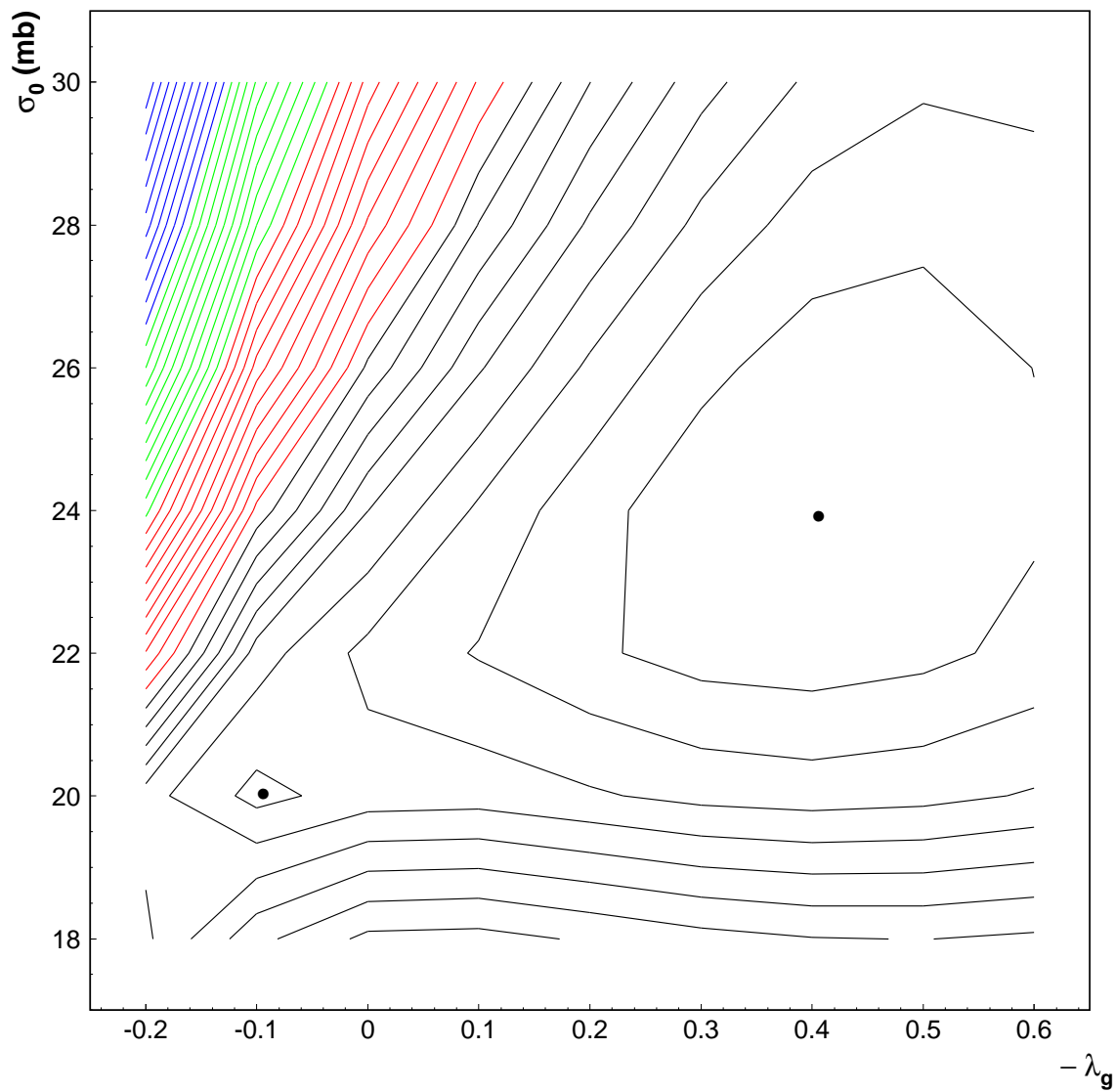,width=17cm}
           }
\vspace*{0.5cm}
\caption{\sl  The lines of constant values of $\chi^2$ in the space
of $(\sigma_0,-\lambda_g)$ in the massless case, $m_q=0$. The two local minima
are indicated by the black dots. The one for $-\lambda_g\simeq 0.4$ corresponds
to FIT 2.
\label{fig:chi2}}
\end{figure}

In a second step of our investigations we relax our  requirement of
staying in the low-$Q^2$ region as close as possible to the original
model. In particular,  we allow the quarks in the $q\bar{q}$ dipole
to become massless. Thus we set $m_q=0$ in the wave
function $\Psi_{T,L}$  and in the kinematic relation (\ref{eq:bjv}).
In the original model, the quark mass was introduced
as an effective parameter for modelling the large-$r$ behaviour of the
photon wave function. The non-zero quark mass allows
to study the photoproduction limit of our model
after the modification  (\ref{eq:bjv}) of the Bjorken-$x$ in the dipole
cross section, therefore, setting this parameter to zero eliminates
this possibility. It allows, however, for  a better description of the current data.
We also  fix the minimal value $\mu_0^2$ of the scale $\mu^2$ in
eq.~(\ref{eq:sighatnew}) to $1~\mbox{\rm GeV}^2$   in the fits
in order to avoid negative gluon density below the input scale $Q_0^2=1~\mbox{\rm GeV}^2$
for the gluon evolution\footnote{The valence-like
gluon density preferred in the massless fit, as described in the text,
becomes negative   below the input scale
due to backward evolution. In this case the dipole
cross section (\ref{eq:sighatnew}) does not saturate at large  $r$.}.
Since the parameters $\sigma_0$ and
$\lambda_g$ are strongly correlated, we have performed a systematic
search  of the best $\chi^2$ on the grid of fixed $(\sigma_0$, $\lambda_g)$.
In each case, the remaining parameters, $A_g$ and $c$, were fitted.
In this way we found two local minima for $\chi^2$, shown in Fig.~\ref{fig:chi2},
for positive value
of $\lambda_g$ in eq.~(\ref{eq:gluon}) leading to  strongly rising gluon density,  and
for negative value of $\lambda_g$, corresponding to valence-like initial gluon.
The latter scenario gives considerable better description
(with $\chi^2/N_{df}=0.97$) than the first
one (with $\chi^2/N_{df}=1.13$).
In the final analysis, after a quantitative estimation of the position of
the best fit in the parameter space using the grid method, we allow
$\sigma_0$ and  $\lambda_g$ to be fitted together with $A_g$ and $c$.
The values of these  parameters for the best fit are given in Table 1 ({\bf Fit 2}).
The corresponding value of the gluon momentum at the input scale equals $84\%$.
The effective slope $\lambda(Q^2)$ from the Fit 2 parameterisation is shown as the dotted line
in Fig.~\ref{fig:6}. As expected, slight  differences  between the two fit scenarios
only appear for  small values of $Q^2$, below $1~\gev^2$.

It is interesting to compare the results of the
{ Fit~1} and { Fit~2} since they
lead to a different picture of the dynamics of
the $\gamma^* p$ interaction. In { Fit~1}, the initial
gluon density, $xg(x,Q^2_0)$, quickly rises with $1/x$, whereas in
{ Fit~2} it even decreases with rising $1/x$. Therefore, in { Fit~1}
the rise of the cross
section with the energy is mainly due to the intrinsic properties of the
initial gluon density, with only slight corrections being due to the
evolution effects at high $Q^2$ values, and considerable damping effects resulting
from saturation at low $Q^2$. In { Fit~2}, the evolution effects are very
strong (note the value of the parameter $C$, which is much higher then in
{ Fit~1}). The small-$x$ rise of the cross section is due solely to the DGLAP evolution
effects with some corrections coming from saturation.

\begin{figure}[p]
\hspace{0.cm}
\centerline{
\epsfig{file=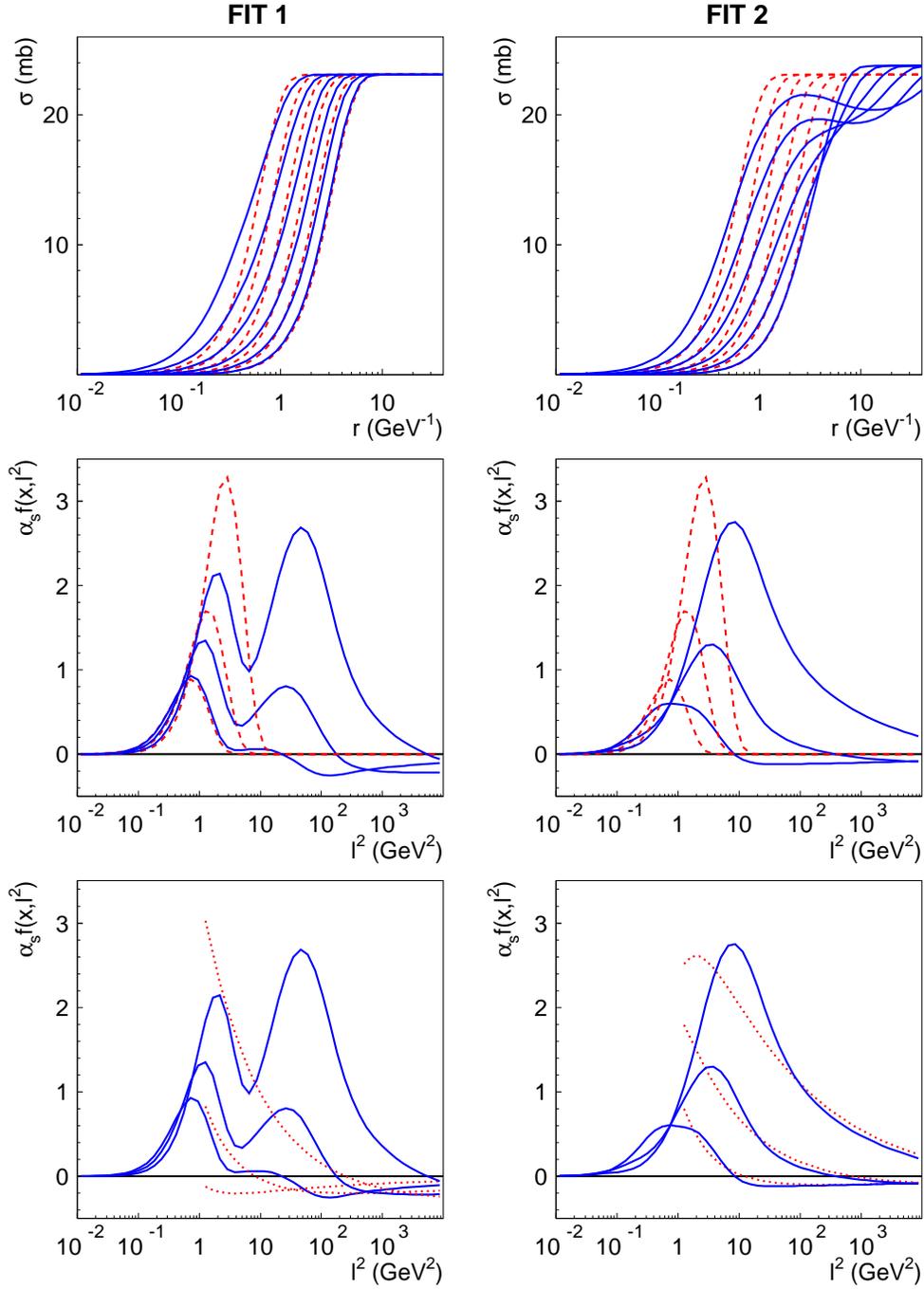,width=15.0cm}
}
\caption{\sl The dipole cross section  $\hat\sigma(x,r)$ (upper row) for
$x=10^{-2}...~10^{-6}$ (from right to left)
and the gluon amplitude $\alpha_s f(x,l^2)$  at {$x=10^{-2}...~10^{-4}$}
(from bottom to top)
for the two fits. The solid lines corespond to the DGLAP improved
model while the dashed lines
describe the saturation model (\ref{eq:sighat}). The dotted lines in the bottow
row show the r.h.s of eq.~(\ref{eq:10}).
}
\label{fig:1a}
\end{figure}

Further insight into the physical picture lying behind the fits can be gained 
by a closer look at $r$-dependence of the dipole cross section 
$\hat\sigma (x,r)$, and a momentum dependence of the related gluon
amplitude, $f(x,l^2)$.
In the first row of Fig.~\ref{fig:1a} we show the dipole cross section
for the two fits.
As we have already discussed, for Fit 1 the DGLAP modification (solid lines) affects mostly the
region of moderately small $r$ ($< 1~\mbox{\rm GeV}^{-1}$). In Fit 2 both
the small  and large $r$  regions are affected. In particular,
the structure close to the saturation region is different
from that in the saturation model (\ref{eq:sighat}) (dashed lines).
The differences  between the models
are particular visible if we  turn to momentum space and
compute the gluon amplitude $\alpha_s f(x,l^2)$ for different values of $x$,
using relation (\ref{eq:4}).
The results are shown in the second row of  Fig.~\ref{fig:1a},
where the full lines  denote the gluon amplitude  from the DGLAP improved model
and the dashed lines correspond to the saturation model (\ref{eq:sighat}).
The small $r$ region of the dipole cross section corresponds to the large $l^2$ region
in the gluon amplitude. The dipole cross section from Fit 1 is translated
into a gluon amplitude with a double-bump structure. Notice that the second bump results
from the DGLAP modification of the small-$r$ part of the dipole cross section.
In Fit 2, however, the  second bump disappears  and the gluon amplitude is similar
but significantly broader than the one corresponding to the saturation model.
Although in Fig.~\ref{fig:1a} the various gluon amplitudes are clearly
distinct, after convolution with the impact factors and turning to the $\gamma^*p$ cross
sections, these differences are becoming much less visible.
It is natural to expect, that some of these differences
should become visible in more exclusive final states. As a first example,
one might think of DIS diffraction. The inclusive diffractive process,
however, is sensitive to the region of small $l^2$ or large $r$ and,
therefore, has only limited value in distinguishing
between the two fit solutions. On the other hand,
other physics processes like jet, charm and
bottom production should be more sensitive to the
behaviour of the unintegrated gluon density at large gluon momenta $l^2$.

Looking at Fig.~\ref{fig:1a}
notice that starting from certain values of $l^2$, the gluon amplitudes become negative. In order
to understand this let us differentiate the relation (\ref{eq:2}) with respect to the large scale
$\mu^2$,
\be
\label{eq:10}
\alpha_s f(x,\mu^2)\,\simeq\,
\frac{\partial}{\partial \ln \mu^2}\,\left\{\alpha_s(\mu^2)\,xg(x,\mu^2)
\right\}
\,=\,\alpha_s(\mu^2)\,\left\{
\frac{\partial xg(x,\mu^2)}{\partial \ln \mu^2}\,-\,\frac{xg(x,\mu^2)}{\ln (\mu^2/\Lambda^2)}
\right\}\,.
\ee
The quantity in the curly brackets in the last equality can become negative, which is shown in
the bottom row of Fig.~\ref{fig:1a} by plotting the r.h.s of the above equation
(dashed lines). In the shown range of $l^2$, eq.~(\ref{eq:10}) is especially well
satisfied for the parameters from Fit 2. For Fit 1 the equality is reached for much larger
(not shown) values of $l^2=\mu^2$.

\begin{figure}[p]
\hspace{0.cm}
\centerline{
\epsfig{file=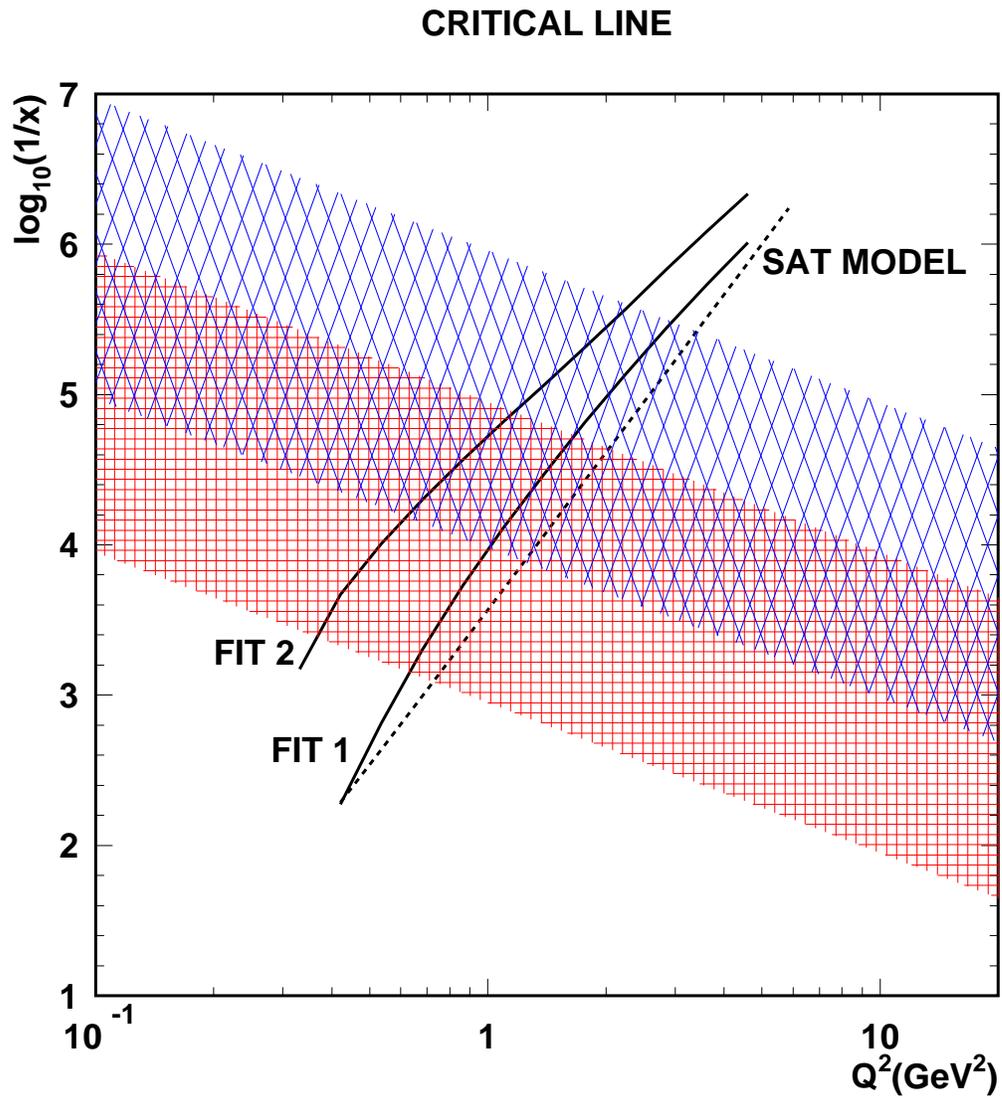,width=15.0cm}
}
\caption{\sl The position of the critical line in the $(x,Q^2)$ plane
in the DGLAP improved model (solid lines) and the original saturation model
(dshed line). The bands indicate acceptance regions for the  colliders
HERA (lower) and future THERA (upper).
}
\label{fig:5}
\end{figure}

In ref.~\cite{GBW1} the critical line in the $(x,Q^2)$ plane was defined which
marks the transition to the saturation region where a new behaviour of the structure
function, $F_2\sim Q^2$, emerges. Near this line,
the characteristic size of the $q\bar{q}$ dipole, $\bar{r}\simeq 2/Q$, equals
the saturation radius $R_0(x)$, see section \ref{sec:1}. In this case the argument
of the exponent in eq.~(\ref{eq:sighat}) equals one,
\be
\label{eq:cr1}
Q^2 R_0^2(x)\,=\,1\,
\ee
and $\hat\sigma(x,\bar{r})\sim \sigma_0$. We adopt the same criterion
for the critical line in the DGLAP improved saturation model (\ref{eq:sighatnew}).
Thus, we have the following condition 
\be
\label{eq:cr2}
\frac{4 \pi^2}{3\,\sigma_0\, Q^2}\,{\alpha_s(\mu^2)\,xg(x,\mu^2)}\,=\,1\,,
\ee
where $\mu^2=C\, Q^2/4+\mu_0^2$.  Eq.~(\ref{eq:cr2})
is an implicit equation for
the critical line $Q^2=Q_s^2(x)$, shown in Fig.~\ref{fig:5}
for the two fits. As expected, for FIT 1 the found critical line
is  not different from that defined in the original
saturation model (dashed line) and
the transition region stays around $1~\mbox{\rm GeV}^2$ in the
HERA kinematics (lower band). For FIT 2 the critical line is situated at lower values
of $Q^2$ (around  $0.5~\mbox{\rm GeV}^2$). It is interesting to note that both fits predict that
in the THERA kinematic range (upper band) the saturation region
lies at $Q^2\approx 2~\mbox{\rm GeV}^2$, which puts the perturbative QCD description
of saturation effects on more solid ground.

\section{Diffraction}

One of the main advantages of dipole models is their straightforward
description of diffractive processes. The generalised optical theorem applied in the
framework of the dipole picture allows to express the cross section for
diffractive $q\bar q$ production in which proton remains intact as
\be
\label{eq:diffcross}
\frac{d \sigma^{\gamma^* p}_{dif}}{dt}_{\mid\, t=0}
\,=\,
\frac{1}{16\pi}\,
\int d^2r \int dz\, \psi^*_{T,L}(Q,r,z)\,
\hat\sigma^2 (x,r)\,
\psi_{T,L}(Q,r,z),
\ee
where $t=\Delta^2$, and $\Delta$ is the four-momentum transfered
into the diffractive system from the proton.
In addition to the contributions of the $q\bar q$ states it is important
to include  the contributions of the $q\bar q g$ final states \cite{BEKW}.

In the phenomenological analysis \cite{GBW2}, the  $q\bar q g$ diffractive amplitude
was computed in the two-gluon exchange approximation  with an additional assumption
of strong ordering  of transverse momenta of the $q\bar{q}$ pair and the gluon.
This allows to treat the  $q\bar q g$ system as a  color octet dipole ($8$$\bar{8}$)
in the transverse coordinate representation.
Compared to  the triplet dipole,
the coupling of two $t$-channel gluons in the singlet state
to the octet dipole carries the relative weight $C_A/C_F=2 N_c^2/(N_c^2-1)$. Thus,
in order to take into account  the repeated exchange of a two gluon system,
the equation (\ref{eq:sighatnew}) for the triplet dipole cross section is modified
for the octet dipole as
\be
\label{eq:sighatmod}
\hat\sigma_{gg} (x,r)\,=\,
{\sigma_0}\,
\left\{
1\,-\,\exp\left(-\frac{C_A}{C_F}\cdot \frac{\pi^2\,r^2\,\alpha_s(\mu^2)\,xg(x,\mu^2)}
{3\,\sigma_0}\right)
\right\}\,.
\ee
The above modification is done in the spirit of  multiple pomeron exchange,
i.e. the term proportional to $({r^2})^n$, resulting
from the expansion of the exponent in (\ref{eq:sighatmod}),
would correspond to $n$ exchanged pomerons
with an appropriate colour factor $(C_A/C_F)^n$. In addition, compared to the
diffractive $q\bar{q}$ production,
the cross section formula for
a diffractive $q\bar{q}g$ system contains an overal factor $(N_c^2-1)/N_c$.

\begin{figure}[hbt]
\vspace{-2.cm}
\epsfig{file=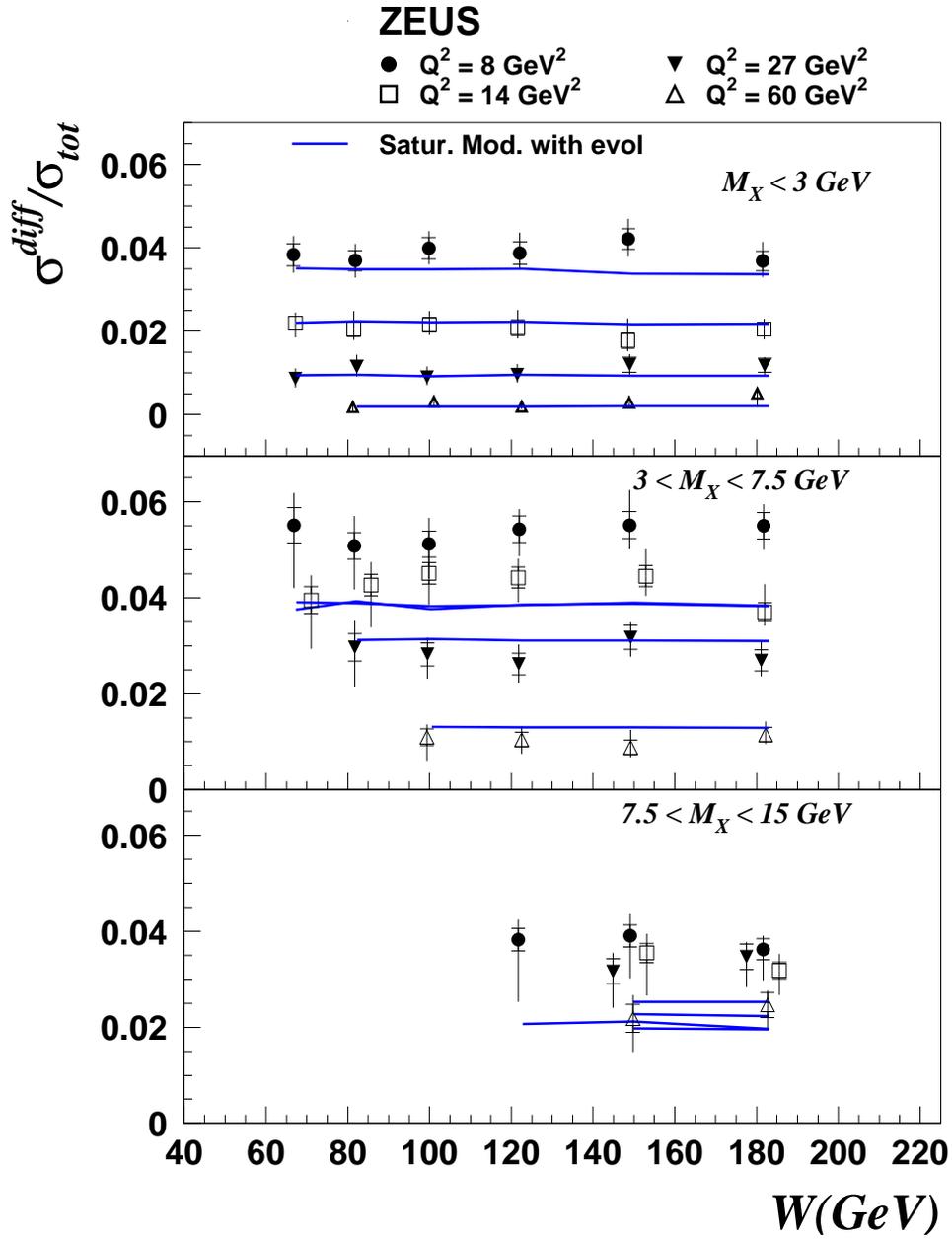,width=15cm}
\caption{\sl The ratio of $\sigma_{diff}/\sigma_{tot}$ versus the $\gamma^*p$ energy $W$.
The data is from ZEUS and the solid lines correspond to the results of the DGLAP improved
model with massless quarks (FIT 2).}
\label{fig:rdiff}
\end{figure}

One of the most important results of the original saturation model was that,
at fixed $Q^2$, the ratio of the inclusive diffractive cross section and
the total $\gamma^* p$ cross section is nearly constant in agreement
with data. This prediction is not changed in the DGLAP improved saturation model
since we modified only the short distance part of the dipole cross section.
Even in the case of Fit~2, the constant ratio is preserved, as shown
in Fig.~\ref{fig:rdiff}, in contrast to the attempts in \cite{LL}. The theoretical
curves in these figures are computed using the Fit 2 results, and the experimental data
are taken from ZEUS \cite{ZEUSDD}.
The results
for the Fit 1 computation differ only slightly from the Fit 2 one.

The diffractive data shown in Fig.~\ref{fig:rdiff} were obtained
without the experimental identification of the forward going proton.
Therefore,
as described in \cite{ZEUSDD}, this data have  a substantial contribution
of the proton dissociation process
which was estimated as $31\pm 15\% $.  To take into account
this contribution,  the prediction of our
model shown in Fig.~\ref{fig:rdiff} were multiplied by a normalisation
factor of $1/(1-0.31) = 1.45$. The agreement of the hight of the predicted
cross section with the data is satifactory only within the relativly large
error of the estimated proton dissociation factor.

We also made a comparison of our predictions  with the recent preliminary diffractive
data in which the forward going proton was identified in the ZEUS Leading
Proton Spectrometer (LPS) \cite{LPS}. Good agreement was found which gives
further support to the dynamical picture of the $\gamma^* p$ interactions
developed in this and previous papers on the saturation model.

\section{Conclusions}
In this letter we have proposed a modification of the saturation model 
which takes into account the QCD DGLAP evolution of the gluon distribution. 
Fitting the parameters of our model
we found a solution that describes the new HERA data on $F_2$
significantly better than the original saturation model,
especially in the region of larger $Q^2$.
The agreement with the DIS diffractive HERA data is also kept.

Somewhat surprisingly, we found another set of parameters which lead to
even better data description
departing from the original saturation model. For this description,  we
set the effective quark mass of the original model equal to zero, and
in our comparison with the HERA data we have disregarded
photoproduction  data points.
We found indications that this solution represents a slightly
different physical picture: the initial gluon density no longer
rises at small $x$, and QCD-evolution plays a much more significant role than
in our first solution. The fact that the effective quark mass of the original 
model has its strongest influence in the limit $Q^2 \to 0$ suggests that 
the large-$r$ behavior of the photon wave function requires further 
considerations.    
 
We have found it useful to discuss the various versions of the saturation 
model not only in the $r$-space but also in momentum space since the latter provides more
direct connection with exclusive final states. As a future step, it will be 
instructive to trace saturation effects in less inclusive cross sections.
      
We consider the modification of the saturation model presented in this paper 
as a first step of a more systematic program.  The success of the original
model indicates that this simple ansatz
contains elements of the 
correct dynamics. Next, we have to analyse this model within QCD
and to find the necessary corrections. With precise HERA data on various 
reactions becoming available 
available, all modifications have to be testet by careful comparisons.

\section*{Acknowledgments}

We thank Jan Kwiecinski and Misha Ryskin for useful discussions.
This research has been supported in part  by
the Polish KBN grant No. 5 P03B 144 20 and
the  Deutsche Forschungsgemeinschaft fellowship.

\newpage

\newpage

\newpage

\newpage

\newpage

\newpage

\end{document}